\documentclass[aps,pra,twocolumn]{revtex4}
\usepackage{amsfonts}
\usepackage{amsmath}
\usepackage{mathrsfs}
\usepackage{amssymb}
\usepackage{graphicx}
\usepackage{bm}

\makeatletter
\renewcommand*\env@matrix[1][\arraystretch]{
  \edef\arraystretch{#1}
  \hskip -\arraycolsep
  \let\@ifnextchar\new@ifnextchar
  \array{*\c@MaxMatrixCols c}}
\makeatother

\begin{document}

\title{Chiral supersolid and dissipative time crystal in Rydberg-dressed Bose--Einstein condensates with Raman-induced
spin--orbit coupling}
\author{Xianghua Su}
\affiliation{Key Laboratory for Microstructural Material Physics of Hebei Province, School of Science, Yanshan University, Qinhuangdao 066004, China}
\author{Xiping Fu}
\affiliation{Key Laboratory for Microstructural Material Physics of Hebei Province, School of Science, Yanshan University, Qinhuangdao 066004, China}
\author{Yang He}
\affiliation{Key Laboratory for Microstructural Material Physics of Hebei Province, School of Science, Yanshan University, Qinhuangdao 066004, China}
\author{Ying Shang}
\affiliation{Key Laboratory for Microstructural Material Physics of Hebei Province, School of Science, Yanshan University, Qinhuangdao 066004, China}
\author{Kaiyuan Ji}
\affiliation{Key Laboratory for Microstructural Material Physics of Hebei Province, School of Science, Yanshan University, Qinhuangdao 066004, China}
\author{Linghua Wen}
\email{linghuawen@ysu.edu.cn}
\affiliation{Key Laboratory for Microstructural Material Physics of Hebei Province, School of Science, Yanshan University, Qinhuangdao 066004, China}
\date{\today }

\begin{abstract}
Spin--orbit coupling (SOC) is one of the crucial factors that affect the chiral symmetry of matter by causing the spatial symmetry breaking of the system.
We find that Raman-induced SOC can induce a chiral supersolid phase with a helical antiskyrmion lattice in balanced Rydberg-dressed two-component Bose--Einstein condensates (BECs) in a harmonic trap by modulating the Raman coupling strength.  This is in stark contrast to the mirror symmetric supersolid phase containing skyrmion--antiskyrmion lattice pair for the case of Rashba SOC. Two ground-state phase diagrams are presented as a function of the Rydberg interaction and the Raman-induced SOC. It is shown that the interplay among Raman-induced SOC, Rydberg interactions, and nonlinear contact interactions favors rich ground-state structures, including half-quantum vortex phase, stripe supersolid phase, toroidal stripe phase with a central Anderson--Toulouse coreless vortex, checkerboard supersolid phase, mirror symmetric supersolid phase, chiral supersolid phase and standing-wave supersolid phase. In addition, the effects of rotation and in-plane quadrupole magnetic field on the ground state of the system are analyzed. In these two cases, the chiral supersolid phase is broken and the ground state tends to form a miscible phase. Furthermore, we demonstrate that when the initial state is a chiral supersolid phase the rotating harmonic trapped system sustains dissipative continuous time crystal by studying the rotational dynamic behaviors of the system.
\end{abstract}

\maketitle


\section{Introduction}
Supersolid is an exotic quantum state of matter possessing simultaneously a superfluid nature and the translational symmetry-breaking feature of solid structures \cite{Leggett,Chester,Leonard}. The initial exploration of supersolids mainly focused on helium, but these studies
of solid $^{4}$He at low temperature ultimately did not find conclusive evidence of supersolidity \cite{Boninsegni,Kim1,Kim2}. Recently, Bose--Einstein condensates (BECs), as an excellent candidate for studying supersolid due to their extremely high purity and full experimental controllability, have attracted a great deal of attention. Relevant experimental and theoretical investigations have shown that the supersolid can be
realized in ultracold atomic BECs by spin--orbit coupling (SOC) \cite{Ketterle}, dipole--dipole interaction (DDI) \cite{Modugno1,Pfau,Ferlaino,Modugno2,Ripley,LChomaz}, optical lattices \cite{Wessel,Landig} or
Rydberg dressing \cite{Henkel1,Cinti,Henkel2,Hsueh1,Zeiher,Seydi}. In a dipolar BEC, the partially attractive nature of the DDI tends to make the system unstable towards a collapse, while the repulsive interaction induced by quantum fluctuations can stabilize the system and lead to the formation of stable quantum droplets \cite{Petrov,Santos,Ferrier-Barbut1,Baillie2}. The quantum droplets may form supersolid under the appropriate conditions \cite{Modugno1,Pfau,Ferlaino,Baillie2,Roccuzzo,XSu}, where the spontaneous periodic density modulation is accompanied by the phase coherence between the droplets. For trapped quasi-one-dimensional (quasi-1D) and quasi-2D dipolar BECs, the dipolar supersolids appear in various spatially-periodic arrangements of droplets, such as straight-line pattern, triangular lattice, square lattice, hexagonal lattice and honeycomb lattice \cite{Ferlaino,Hertkorn,Young,Hertkorn2}. The physical systems mentioned above are hard-core systems, where the interatomic contact interaction or long-range DDI is essentially a hard-core interaction. By contrast, the Rydberg interactions (van der Waals interactions) in Rydberg-dressed BECs are soft-core isotropic long-range interactions. The strong van der Waals interactions are expected to break the spatial translational symmetry in multiple directions and induce the formation of 2D and 3D supersolids \cite{Henkel1,Hsueh1}.

On the other hand, SOC is an interaction between the spin and the momentum of a quantum particle, which breaks the spatial symmetry and the Galilean invariance. The ultracold Bose gases with SOC favor many novel quantum states \cite{Spielman1,Zhai,Spielman2}, such as plane-wave phase \cite{Zhai}, stripe phase \cite{YZhang,HWang}, supersolid phase \cite{Ketterle,WHan}, lattice phase \cite{Sinha}, various vortex states \cite{XQXu,Radic,Aftalion}, soliton excitation \cite{Xu,Sakaguchi,HuangC,MalomedBA,ZhaoY,ZhaoD} and
skyrmion \cite{HHu}. For Rashba spin--orbit-coupled spin-1/2 BEC in a harmonic trap, the ground state presents as a plane wave phase (miscible interactions, i.e. the interactions for phase mixing) or a stripe phase (immiscible interactions, i.e. the interactions for phase separation), depending on the competition between the intra- and interspecies interactions \cite{Zhai,CWang,Gui,Baizakov}. Recent studies show that the combined effects of SOC and nonlocal soft-core long-range interactions can lead to rich quantum phases \cite{Lu,Han,Lyu,Zhang}. In particular, the Rashba SOC or Dresselhaus SOC is predicted to achieve the symmetry breaking of chirality and induce a chiral supersolid in uniform two-component BECs with imbalanced Rydberg interactions \cite{Han,Zhang}. In the latest years, increasing attention has been paid to the study of chiral matter, which provides new ideas for the design of new materials \cite{Kallin,Yang,Szasz,LWang}.

In addition, 1D, 2D and 3D Raman-induced SOC in atomic BECs have been experimentally realized by Raman coupling that flips atomic pseudospins and transfers linear momentum simultaneously \cite{Spielman1,Pan1,Pan2}. Although Rashba SOC has also been experimentally realized in BECs \cite{Valdes}, achieving Rashba SOC is generally much more difficult than achieving Raman-induced SOC. Furthermore, relevant investigations have shown significant differences between the Raman-induced SOC effects and the Rashba SOC effects. For instance, for a binary BEC with Rashba SOC, a stripe phase can exist in immiscible interactions \cite{Zhai,CWang,Gui}. In contrast, in a binary BEC with Raman-induced SOC, the stripe phase may occur in the miscible regime \cite{Ketterle,Spielman1,YLi}. Up to date, current studies on Raman-type spin--orbit-coupled BECs have primarily focused on 1D Raman-induced SOC. The change in Raman coupling strength can break spatial symmetry \cite{Zhai}, but it is not yet clear whether it will lead to chiral supersolid and other novel quantum phases in Rydberg-dressed BECs with 2D Raman-induced SOC. Moreover, spontaneous time symmetry breaking may induce an exotic dynamical phase, the time crystal \cite{Wilczek,Sacha,HuangB,Rovny,Hemmerich}. Previous studies have shown that time crystal can be observed in a variety of physical systems, such as Floquet many-body localized (Floquet-MBL) system \cite{Monroe}, dipolar many-body system \cite{Choi},
dissipative atom-cavity BEC system \cite{Kongkhambut}, and Rydberg gas \cite{You}. Whether the time-reversal symmetry breaking caused by SOC can result in the formation of time crystals is an issue worth exploring.

In this paper, we consider quasi-2D two-component BECs with Raman-induced SOC and balanced Rydberg interactions in a harmonic trap. Two ground-state phase diagrams are given. Due to the combined effects of Raman-induced SOC, balanced Rydberg interactions and immiscible contact interactions, there are much richer quantum phases than those in the two-component homogeneous BECs with Rashba (Dresselhaus) SOC, imbalanced Rydberg interactions and miscible contact interactions \cite{Han}. In particular, we demonstrate that Raman-induced SOC can lead to a greater variety of stripe phases and supersolid phases. Specifically, the quantum phases include half-quantum vortex phase, stripe supersolid phase, toroidal stripe phase with a central Anderson-Toulouse coreless vortex, checkerboard supersolid phase, standing-wave supersolid phase, mirror-symmetric supersolid phase with skyrmion-antiskyrmion lattice pair, and chiral supersolid phase with a helical antiskyrmion lattice. And these quantum phases can be achieved by modulating the Rydberg interaction strength, SOC strength, and Raman coupling strength. Additionally, it is shown that rotation and in-plane gradient magnetic field can give rise to the destruction of the chiral supersolid phase and transform the ground state of the system into a miscible phase. Moreover, we investigate the rotating dynamics of the system by using a phenomenological dissipation model, and find that the system supports dissipative continuous time crystals.

This paper is organized as follows. In Sec. II, we formulate the theoretical model and methods. In Sec. III, we first study the ground-state structures of the system, and then discuss the effects of rotation and in-plane quadrupole magnetic field on the ground state of the system. Next, we investigate the rotational dynamic behaviors of the system. The main conclusions of the paper are summarized in the last section.
\section{Theoretical model}
We consider a quasi-2D spin-1/2 BEC with 2D Raman-induced SOC and Rydberg interactions in a harmonic trap. Under the mean-field approximation, the energy functional of the system is given by \cite{Henkel1,Hsueh1,Zhai,Han,SYang,Sachdeva,YuZhang,WangY}
\begin{align}
E=& \int d\mathbf{r}\psi^{\dag} \Big[-\frac{\hbar ^{2}}{2m}\nabla
^{2}+V(\mathbf{r})+\upsilon _{so}\Big]\psi \notag \\
&+\frac{1}{2}\int d\mathbf{r}\sum_{j,l=1,2}g_{jl}\psi^{*}_{j}(\mathbf{r})\psi^{*}_{l}(\mathbf{r})\psi_{l}(\mathbf{r})\psi_{j}(\mathbf{r})  \notag \\
&+\frac{1}{2}\int d\mathbf{r}d\mathbf{r^{\prime}}\sum_{j,l=1,2}\psi^{*}_{j}(\mathbf{r})\psi^{*}_{l}(\mathbf{r^{\prime}})  \notag \\
&\times U_{jl}\left(\mathbf{r-r^{\prime }}\right)\psi_{l}(\mathbf{r}^{\prime})\psi_{j}(\mathbf{r}).  \label{1}
\end{align}%
Here $m$ is the atomic mass, $\psi =\left[ \psi _{1}\left( \mathbf{r}%
\right) ,\psi _{2}\left( \mathbf{r}\right) \right] ^{T}$ with $\mathbf{r=}%
\left( x,y\right) $, $\psi _{j}\left( j=1,2\right) $ is the component
wave function, with 1 and 2 corresponding to spin-up (component 1) and
spin-down (component 2), respectively. The system satisfies the normalization condition $%
\int d\mathbf{r}( |\psi _{1}|^{2}+|\psi _{2}|^{2}) =N$. $V(\mathbf{r})=m\omega _{\bot }^{2}(x^{2}+y^{2})/2$ is the 2D external trapping
potential with the radial trap frequency $\omega _{\bot }$. We assume that the system is tightly confined by a harmonic
trap in the $z$ direction to form a quasi-2D system with frequency $\omega
_{z}$ and characteristic length $a_{z}=\sqrt{\hbar /m\omega _{z}}$. The 2D Raman-induced SOC reads \cite{SYang,Sachdeva,YuZhang,SalasnichL}
\begin{equation}
\upsilon _{so}=-i\hbar \kappa \left( \sigma _{x}\partial _{x}+\sigma
_{y}\partial _{y}\right) +\frac{\Omega _{R}}{2}\sigma _{z}-\frac{\delta }{2}%
\sigma _{x},  \label{2}
\end{equation}%
where $\bm{\sigma}=(\sigma _{x},\sigma _{y},\sigma _{z})$ are the Pauli
matrices, $\kappa $ is the SOC strength, $\Omega _{R}$ is the Raman coupling
strength, and $\delta $ is the Raman laser detuning, representing the deviation between the frequency difference of two Raman lasers and the transition frequency of two atomic spin states \cite{Spielman1,Lu}. The coefficients $g_{jj}=2\sqrt{2\pi }a_{j}\hbar ^{2}/ma_{z} (j=1,2)$ and $%
g_{12}=g_{21}=2\sqrt{2\pi }a_{12}\hbar ^{2}/ma_{z}$ represent the intra- and
interspecies coupling strengths, where $a_{j}$ and $a_{12}$ are the $s$-wave
scattering lengths between intra- and intercomponent atoms.  In the mean time, the soft-core long-range
interaction potential, i.e., the nonlocal Rydberg interaction potential, can be expressed as \cite{Henkel1,Hsueh1}
\begin{equation}
U_{jl}\left( \mathbf{r-r^{\prime }}\right) =\frac{\widetilde{C}_{6}^{jl}}{R_{c}^{6}+\left\vert \mathbf{%
r-r^{\prime }}\right\vert ^{6}}.  \label{3}
\end{equation}%
Here we assume that the system has equilibrium Rydberg
interactions for $\widetilde{C}_{6}^{11}=\widetilde{C}_{6}^{12}=\widetilde{C}%
_{6}^{21}=\widetilde{C}_{6}^{22}=\widetilde{C}_{6}$. $\widetilde{C}%
_{6}=\left( \Omega /2\Delta \right) ^{4}C_{6}/\sqrt{2\pi }a_{z}$, where $\Omega$ and $\Delta $ are the Rabi frequency and detuning of the Rydberg-dressing laser which creates a nonlocal interaction between Rydberg-dressed atoms by off-resonant coupling the electronic ground states of the atoms to Rydberg states \cite{Henkel1,Zeiher,Lu,Heidemann}. $R_{c}=\left( C_{6}/2\hbar \Delta
\right) ^{1/6}$ is blockade radius, and $C_{6}$ is the strength of the van
der Waals interaction between Rydberg atoms \cite{Henkel1}.

For the sake of numerical simulation and calculation, we introduce the notations $\widetilde{t}=t/\tau$, $\widetilde{\mathbf{r}}=\mathbf{r}/R_{c}$, $\omega_{c}=\omega_{\perp}\tau$, $\widetilde{\Omega}_{R}=\Omega_{R}\tau/\hbar$, $\widetilde{\delta}=\delta\tau/\hbar$, $\widetilde{\kappa}=\kappa \tau/R_{c}$, $\widetilde{\psi}_{j}=\sqrt{R_{c}^3/N}\psi _{j}(j=1,2)$, $\widetilde{U}_{jl}\left(\mathbf{r-r^{\prime}}\right)=mN\widetilde{C}_{6}^{jl}/{\hbar}^2{R_{c}^{6}}(1+\left\vert \mathbf{\widetilde{r}-\widetilde{r}^{\prime}}\right\vert^{6})$, and $\tau=R_{c}^2m/\hbar$. Here the blockade radius $R_{c}$ and $\tau$ are chosen as the
length and time scales.  Then we obtain the dimensionless 2D coupled GP equations
\begin{align}
i\frac{\partial \psi _{1}\left( \mathbf{r}\right) }{\partial t}=& \Big[-\frac{1}{2}\nabla ^{2}+V(\mathbf{r
})+\beta _{11}|\psi _{1}\left( \mathbf{r}\right) |^{2}+\beta _{12}|\psi _{2}\left(\mathbf{r}\right) |^{2}  \notag \\
&+\frac{\Omega _{R}}{2}+\int U_{11}\left( \mathbf{r}-\mathbf{r}^{\prime }\right)
|\psi _{1}(\mathbf{r}^{\prime })|^{2}d\mathbf{r}^{\prime } \notag \\
&+\int U_{12}\left( \mathbf{r}-\mathbf{r}^{\prime }\right) |\psi _{2}(\mathbf{r}^{\prime })|^{2}d\mathbf{r}^{\prime }\Big]\psi _{1}\left( \mathbf{r}\right)  \notag \\
& -\kappa \left(i\partial _{x}+\partial _{y}\right)\psi _{2}\left( \mathbf{r}\right)-\frac{\delta }{2}\psi _{2}\left( \mathbf{r}\right), \label{4} \\
i\frac{\partial \psi _{2}\left( \mathbf{r}\right) }{\partial t}=& \Big[-\frac{1}{2}\nabla ^{2}+V(\mathbf{r
})+\beta _{22}|\psi _{2}\left( \mathbf{r}\right) |^{2}+\beta _{21}|\psi _{1}\left(\mathbf{r}\right) |^{2}  \notag \\
&-\frac{\Omega _{R}}{2}+\int U_{22}\left( \mathbf{r}-\mathbf{r}^{\prime }\right)
|\psi _{2}(\mathbf{r}^{\prime })|^{2}d\mathbf{r}^{\prime } \notag \\
&+\int U_{21}\left( \mathbf{r}-\mathbf{r}^{\prime }\right) |\psi _{1}(\mathbf{r}^{\prime })|^{2}d\mathbf{r}^{\prime }\Big]\psi _{2}\left( \mathbf{r}\right)  \notag \\
& -\kappa \left(i\partial _{x}-\partial _{y}\right)\psi _{1}\left( \mathbf{r}\right)-\frac{\delta }{2}\psi _{1}\left( \mathbf{r}\right),  \label{5}
\end{align}%
where the tildes are omitted for brevity. The dimensionless 2D harmonic potential can be written as $V(\mathbf{r})=\omega _{c}^{2}(x^{2}+y^{2})/2$ with the strength of the radius potential $\omega _{c}$. The dimensionless intra- and interspecies interaction strengths read as $\beta _{jj}=2\sqrt{%
2\pi }a_{j}N/R_{c}a_{z}(j=1,2)$ and $\beta _{12}=\beta _{21}=2\sqrt{2\pi }%
a_{12}N/R_{c}a_{z}$. For convenience, we rescale the soft-core long-range Rydberg interaction strength as $mN%
\widetilde{C}_{6}/\hbar ^{2}R_{c}^{6}$ and denote it as $\widetilde{C}_{6}$.

To further explore the topological properties of the system, we adopt a nonlinear Sigma model \cite{Kasamatsu2,Aftalion} and introduce a normalized
complex-valued spinor $\bm{\chi }=[\chi _{1},\chi _{2}]^{T}$ with $|\chi
_{1}|^{2}+|\chi _{2}|^{2}=1$. The corresponding two-component wave functions are given by
$\psi _{1}=\sqrt{\rho }\chi _{1}$ and $\psi _{2}=\sqrt{%
\rho }\chi _{2}$, where $\rho =|\psi
_{1}|^{2}+|\psi _{2}|^{2}$ is the total density of the system. The spin density is defined as $\mathbf{S}=%
\overline{\bm{\chi }}\bm{\sigma}\bm{\chi}$, and the components of $\mathbf{S}
$ are written as
\begin{eqnarray}
S_{x} &=&\chi _{1}^{\ast }\chi _{2}+\chi _{2}^{\ast }\chi _{1},  \label{Sx}
\\
S_{y} &=&i(\chi _{2}^{\ast }\chi _{1}-\chi _{1}^{\ast }\chi _{2}),
\label{Sy} \\
S_{z} &=&|\chi _{1}|^{2}-|\chi _{2}|^{2},  \label{Sz}
\end{eqnarray}%
with $|\mathbf{S}|^{2}=S_{x}^{2}+S_{y}^{2}+S_{z}^{2}=1$. The spacial distribution of the topological structure of the system can be well characterized by the topological
charge density%
\begin{equation}
q(\mathbf{r})=\frac{1}{4\pi }\mathbf{S}\cdot \left( \frac{\partial \mathbf{S}%
}{\partial x}\times \frac{\partial \mathbf{S}}{\partial y}\right),
\label{TopologicalChargeDensity}
\end{equation}%
and the topological charge $Q$ is given by
\begin{equation}
Q=\int q(\mathbf{r})dxdy.  \label{TopologicalCharge}
\end{equation}
\section{Results and discussion}

\subsection{Chiral supersolid and phase diagram}
By numerically solving the 2D coupled GP equations (\ref{4})-(\ref{5}) and minimizing the GP energy functional of the system with the imaginary-time propagation method and the time-splitting Fourier pseudospectral method, we can obtain the ground state of the system. We consider $^{87}$Rb BECs with the number of atoms $N\sim1\times 10^{4}$ in a harmonic trap with the trapping frequencies $(\omega_{\bot}, \omega_{z})=2\pi\times (20,200)$ Hz, where $\left\vert F=1,m_f=-1\right\rangle $ and $\left\vert F=1,m_f=0\right\rangle $ spin states of $^{87}$Rb atoms correspond to spin-up and spin-down components, respectively. In our numerical calculations, we have used a $512\times512$ grid size with $\Delta x = \Delta y = 0.02$ for both $x$ and $y$ (units $R_{c}$) and with time step $\Delta t = 0.01$ (units $\tau$). For convenience, we fix the Rydberg blockade radius $R_{c}=1.5$, $\omega _{c}=5$ and $\delta=0$ (Raman resonance case, that is, the frequency difference between two Raman beams is equal to the transition frequency between two atomic spin states), and assume that the intra- and interspecies interaction strengths are $\beta _{11}=\beta_{22}=100$ and $\beta _{12}=200$ (immiscible contact interactions), respectively. The $s$-wave scattering lengths between intra- and intercomponent atoms are $a_{1}=a_{2}=100a_B$ and $a_{12}=200a_B$ with the Bohr radius $a_B$, respectively. The long-range Rydberg interaction can be realized by the Rydberg dressing technique \cite{Henkel1,Zeiher,Heidemann}, and its strength can be changed within a wide range by tuning the two-photon Rabi frequency and detuning of the Rydberg dressing laser. The SOC can be realized by modulating the Raman laser dressing \cite{Spielman1,Spielman2,Pan1,Pan2,Valdes}. Here, the SOC strength $\kappa = 0.1-8$ can be tuned by changing the Raman laser wavelengths in the range from $16.23$ $\mu$m to $202.91$ nm. The dimensionless Raman coupling strength $\Omega _{R}= 0-20$ used in the simulation can be attained by tuning the Raman laser strength in the range of $2\pi \hbar \times (0-80)$ Hz.

In the presence of Rashba SOC or Dresselhaus SOC, it has been reported that the imbalanced intra- and intercomponent Rydberg interactions can induce a chiral supersolid (CSS) phase in homogeneous two-component BECs with miscible contact interactions \cite{Han}. In the CSS phase, the chiral symmetry of the system is broken and the periodic density modulation is spontaneously formed. In the present work, we focus on whether Raman-induced SOC can prompt the formation of similar chiral supersolid phase or other peculiar quantum phases in the balanced intra- and intercomponent Rydberg-dressed BECs. To highlight the effect of Raman coupling, we first consider the case of Rashba SOC ($\Omega _{R}=0$, i.e., without Raman coupling). Experimentally the Rashba SOC can be created by applying inhomogeneous magnetic field \cite{Zhai,Han,BMAnderson}. Under the conditions of equilibrium Rydberg interactions and harmonic trap, the typical density distributions, phase distributions and spin texture of the system are shown in Fig. 1(a). Evidently, the system exhibits a phase separated and periodically density modulated structure and features a mirror symmetry. At the same time, there exists a singular topological structure, where the visible vortices (clockwise rotation) are generated in the upper half space of component 1 while the visible antivortices (anticlockwise rotation) are created in the lower half space of component 2 (see the density and phase distributions in Fig. 1(a)). According to the distribution of the visible vortices, it can be seen that the superfluid is a local current. Therefore, we can call the phase as mirror symmetric supersolid (MSSS) phase. From the spin texture in Fig. 1(a), there is a skyrmion lattice in the $y<0$ region, with the local topological charge of each skyrmion being $Q=1$ \cite{Skyrme,Mermin}, and there is an antiskyrmion lattice in the $y>0$ region, with the local topological charge of each antiskyrmion being $Q=-1$. The skyrmion lattice and the antiskyrmion lattice are symmetrically distributed about the $y=0$ axis, forming skyrmion-antiskyrmion lattice pair. In addition, a spin Neel domain wall is generated along the $y=0$ direction. Physically, the skyrmions in the spin textures of the binary BEC are associated with the vortex structures in the component density distributions, and the particle density must satisfy the continuity condition due to quantum fluid nature of the condensates. Moreover, the spin vector for a skyrmion covers the whole unit sphere of spin space, which is due to the fact that the topological charge is $Q=1$. In other words, for a specific unit cell, a skyrmion means that if the spin-density components $S_x$ and $S_y$ can vary from -1 to 1, then the spin-density component $S_z$ also varies from -1 to 1. The antiskyrmion is the inverse scenario of the skyrmion, i.e., $Q=-1$.
\begin{figure}[tbp]
\centerline{\includegraphics*[width=8cm]{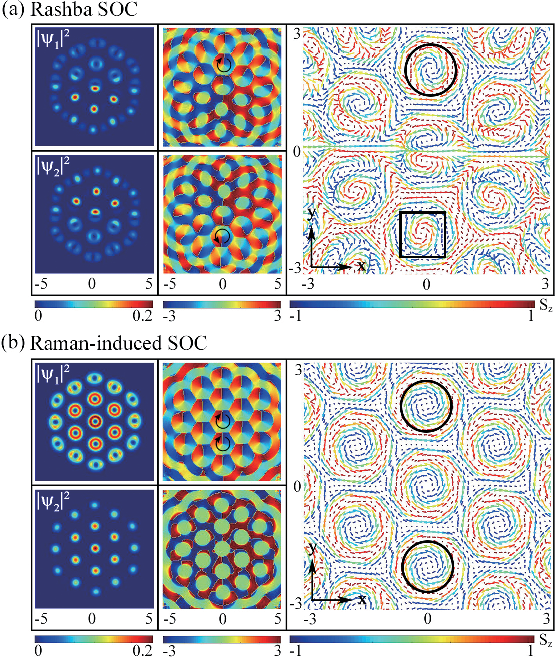}}
\caption{(Color online) Typical density distribution (left), phase distribution (middle) and
spin texture (right) of the system for (a) $\widetilde{C}_{6}=1500$, $\protect\kappa%
=4$, $\Omega _{R}=0$ and (b) $\widetilde{C}_{6}=1500$, $\protect\kappa=4$, $%
\Omega _{R}=10$. The arrows in the spin texture represent the transverse
spin vector ($S_{x}$, $S_{y}$) and the color of each arrow indicates the
magnitude of $S_{z}$. The square and circle in the spin texture denote a skyrmion and an antiskyrmion, respectively. Here each panel is a square, and the range and scale of the vertical axis are the same as those of the horizontal axis. The unit length is $R_{c}$.}
\label{Fig1}
\end{figure}

Next, we study the case of Raman-induced SOC ($\Omega _{R}\neq0$). Intriguingly, we obtain a chiral supersolid phase by applying the Raman-induced SOC and the equilibrium Rydberg interactions as shown in Fig. 1(b). By comparison, the two spin components in this quantum phase are separated along the radial direction in each unit cell possessing a clockwise circulation, where component 2 is located in the center and surrounded by component 1. In the meantime, the antivortices in component 2 disappear, while all the phase defects in component 1 become visible vortices and constitute a triangular vortex lattice containing local and global circulating particle currents. Obviously, the chiral symmetry of the system is broken. Furthermore, our computation results demonstrate that the local topological defects in the spin texture in Fig. 1(b) are helical antiskyrmions \cite{Nagaosa} with topological charge $Q=-1$, thus the spin texture in Fig. 1(b) is an exotic helical antiskyrmion lattice, where the spin Neel domain wall spontaneously vanishes. These results indicate that the Raman laser leads to the reversal of certain local spin structures. Compared with Ref. \cite{Han}, here the novel chiral supersolid phase with helical antiskyrmion lattice can be created in two-component BECs with Raman-induced SOC and balanced Rydberg interactions in a harmonic trap, by adjusting the Raman coupling strength. In this sense, the present system is approach to an actual harmonic-trapped BEC system and may be relatively easy to achieve.
\begin{figure*}[htbp]
\centering
\centerline{\includegraphics*[width=14cm]{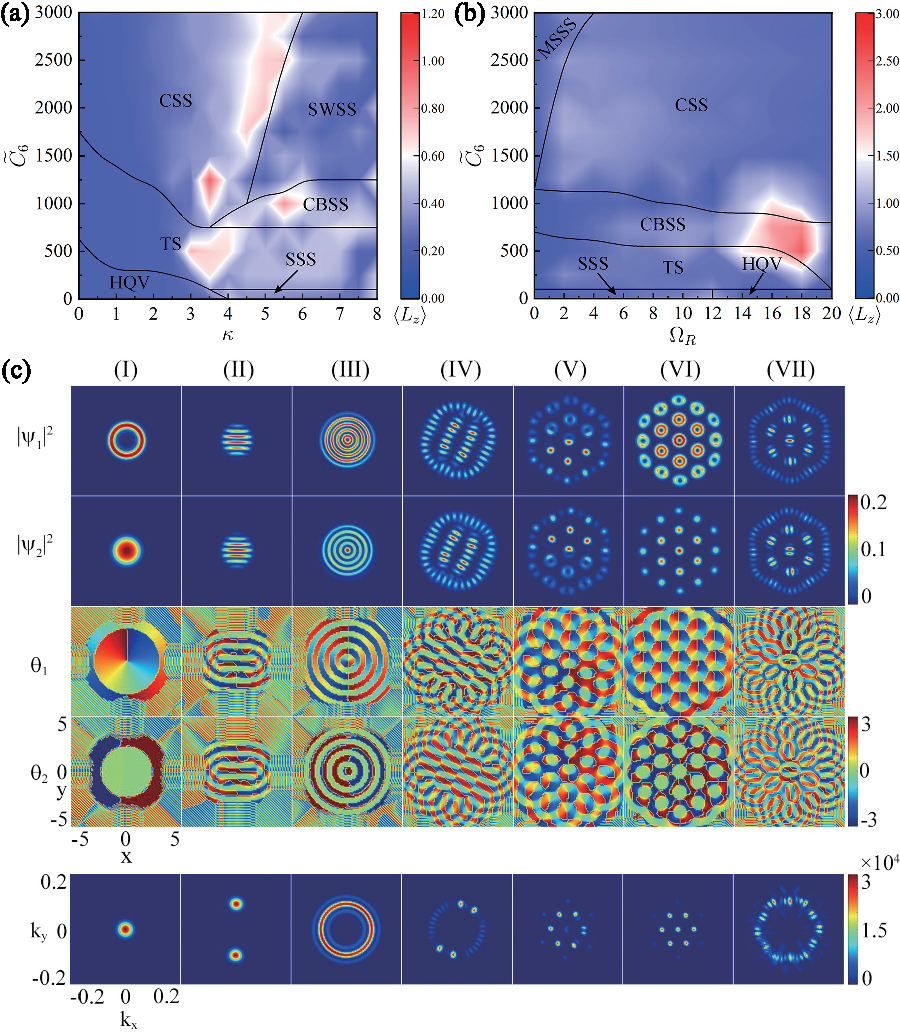}}
\caption{(Color online) (a) Ground-state phase diagram as the function of the SOC strength $\kappa$ and the Rydberg interaction strength $\widetilde{C}_{6}$ for
 two-component BECs with Raman-induced SOC and balanced Rydberg interactions in a harmonic trap, where the Raman coupling strength $\Omega _{R}=10$. (b) Ground-state phase diagram with respect to $\Omega _{R}$ and  $\widetilde{C}_{6}$, where $\kappa=4$. The background color in (a) and (b)
indicates the magnitude of $\left\langle L_{z}\right\rangle $. (c) The
first four rows denote typical density distributions and phase distributions of various ground-state phases, where  (\uppercase\expandafter{\romannumeral1})-(\uppercase\expandafter{\romannumeral7}) correspond to the half-quantum vortex (HQV) phase, stripes supersolid (SSS) phase, toroidal stripe (TS) phase with a central Anderson-Toulouse coreless vortex, checkerboard supersolid (CBSS) phase, mirror-symmetric supersolid (MSSS) phase with skyrmion-antiskyrmion lattice pair, chiral supersolid (CSS) phase with a helical antiskyrmion lattice, and standing-wave supersolid (SWSS) phase, respectively. The last row corresponds to the momentum distribution of the system. The relevant parameters are (\uppercase\expandafter{%
\romannumeral1}) $\widetilde{C}_{6}=10$, $\protect\kappa=2$, $\Omega _{R}=10$%
, (\uppercase\expandafter{\romannumeral2}) $\widetilde{C}_{6}=10$, $\protect%
\kappa=6$, $\Omega _{R}=10$, (\uppercase\expandafter{\romannumeral3}) $%
\widetilde{C}_{6}=250$, $\protect\kappa=6$, $\Omega _{R}=10$, (\uppercase
\expandafter{\romannumeral4}) $\widetilde{C}_{6}=1000$, $\protect\kappa=6$, $%
\Omega _{R}=10$, (\uppercase\expandafter{\romannumeral5}) $\widetilde{C}%
_{6}=1500$, $\protect\kappa=4$, $\Omega _{R}=0$, (\uppercase
\expandafter{\romannumeral6}) $\widetilde{C}_{6}=1500$, $\protect\kappa=4$, $%
\Omega _{R}=10$, and (\uppercase\expandafter{\romannumeral7}) $\widetilde{C}%
_{6}=1500$, $\protect\kappa=7$, $\Omega _{R}=10$. The unit length in the first to fourth rows of Fig. 2(c) is $R_{c}$.}
\label{Fig2}
\end{figure*}

In order to further elucidate the ground-state properties of the system, we provide two ground-state phase diagrams with respect to $\widetilde{C}_{6}$ and $\kappa $ and with respect to $\widetilde{C}_{6}$ and $\Omega _{R}$ as shown in Figs. 2(a) and 2(b), respectively. The typical density distributions, phase distributions and momentum distributions of various ground-state phases are given in Fig. 2(c). From Fig. 2(a), for the weak Rydberg interaction $\widetilde{C}_{6}$ with fixed Raman coupling strength $\Omega _{R}$, the system sustains half-quantum vortex (HQV) phase and stripe supersolid (SSS) phase \cite{Ketterle}, depending on the SOC strength. With the increase of SOC strength, the ground state of the system changes from the HQV phase to the SSS phase. In the HQV phase, it can be seen from the momentum distribution that the atoms are essentially condensed at zero momentum (see Fig. 2(c)(I)). Unlike the HQV phase, the momentum distribution of the SSS phase exhibits two discrete high-density points (see Fig. 2(c)(II)). This indicates that the atoms in the SSS phase are principally condensed at two finite momenta. However, for fixed SOC strength but stronger Rydberg interaction, the two component densities form spatially separated and multiple concentric layered toroidal stripes, and the central vortex core of component 1 is filled by the nonrotating component 2 as shown in Fig. 2(c)(III). We may call this phase as toroidal stripe (TS) phase with a central Anderson-Toulouse coreless vortex \cite{Anderson}. At the same time, the momentum distribution displays an obvious high-density ring and two low-density rings. For the case of strong SOC, with the further increase of the Rydberg interaction strength, the ground state of the system evolves from the TS phase into the checkerboard supersolid (CBSS) phase (Fig. 2(c)(IV)). The density peaks form a regular checkerboard pattern, and many hidden vortex-antivortex pairs are generated in each component \cite{LWen,LWen1,LWen2}. Previous studies have shown that there are three basic types of vortices in cold atom physics: visible vortices, ghost vortices and hidden vortices \cite{LWen,LWen1,LWen2,Fetter,Kasamatsu,WenLH}. The visible vortices represent the common quantized vortices, which are visible in both density and phase distributions and contribute to the angular momentum and energy of the system \cite{Fetter}. In contrast, ghost vortices appear as phase singularities in the phase distribution, whereas they are invisible in the density distribution and do not carry angular momentum and energy \cite{Kasamatsu}. For hidden vortices, they are visible in the phase distribution while invisible in the density distribution like ghost vortices, but they carry angular momentum and energy \cite{LWen,LWen1,LWen2,WenLH}. Only by accounting for hidden vortices can the established Feynman rule be satisfied.
The corresponding momentum distribution in Fig. 2(c)(IV) is focused on four high density points and some low density points
along a ring. Whereas for the case of weak (or relatively weak) SOC and strong Rydberg interaction, the system tends to form the chiral supersolid (CSS) phase with a helical antiskyrmion lattice (see Fig. 1(b), Fig. 2(a) and Fig. 2(c)(VI)).

When both the Rydberg interaction and the SOC are strong, the CSS phase transforms into the standing-wave supersolid (SWSS) phase as shown in Fig. 2(a) and Fig. 2(c)(VII), where the momentum distribution becomes a ring structure composed of stripe density standing waves. Physically, for the SSS phase, CBSS phase and SWSS phase, the translation symmetry of the system is broken due to the strong SOC, resulting in the formation of a supersolid crystal structure of density modulation.

In the phase diagram as the function of $\widetilde{C}_{6}$ and $\Omega _{R}$, we find that in the region of strong Rydberg interaction ($\widetilde{C}_{6}\gtrsim 1000$) and weak Raman coupling, the system supports the MSSS phase with skyrmion-antiskyrmion lattice pair (see Fig. 2(b)and Fig. 2(c)(V)). With the increase of Raman coupling strength, the ground-state phase of the system changes from the MSSS phase to the CSS phase with a helical antiskyrmion lattice. Similarly, there also exist HQV phase, SSS phase, TS phase, and CBSS phase in the phase diagram of Fig. 2(b).

Moreover, the average orbital angular momentum (i.e., canonical angular momentum) per atom, $\left\langle L_{z}\right\rangle
=\sum_{j=1,2}\int d\mathbf{r} \psi _{j} ^{*}\left( xp_{y}-yp_{x}\right) \psi _{j}$ \cite{XQXu}, can also show the influence of the SOC, Rydberg interaction, and Raman coupling on the ground-state structure and phase transition of the system. The dependence of $\left\langle
L_{z}\right\rangle $ on $\kappa $, $\widetilde{C}_{6}$, and $\Omega _{R}$ is displayed in Figs. 2(a) and 2(b), which indicates that there is no simple linear relationship, but a complex correlation associated with the specific ground-state configurations. For instance, the strong SOC overall leads to more angular momentum in the system due to the interaction between the spin and the momentum except for the SWSS phase (see Fig. 2(a)). Near the boundary between the CSS phase and the SWSS phase in the region of strong SOC and strong Rydberg interaction, there is a significant sudden change in $\left\langle L_{z}\right\rangle$. The physical reason is that in the CSS phase there are a large number of
visible vortices carrying evident angular momentum while in the SWSS phase the system forms multiple localized stripe structures with almost no phase defects. However, for the CSS phase, $\left\langle L_{z}\right\rangle $ decreases slightly as the Raman coupling strength $\Omega _{R}$ increases (see Fig. 2(b)). This behavior is due to the presence of a small amount of hidden vortices \cite{LWen,LWen1,LWen2,WenLH} in component 2 for the case of small Raman coupling strength. In short, the comprehensive competition of Rydberg interaction, SOC, and Raman coupling leads to complicated changes in the orbital angular momentum.

As mentioned above, Raman-induced SOC can cause the ground state of the system to exhibit chiral supersolid phase and standing-wave supersolid phase. What's more, this system also sustains novel quantum phases that have not been reported in Ref. \cite{Han}, including half quantum vortex phase, stripe supersolid phase, toroidal stripe phase, checkerboard supersolid phase and mirror-symmetric supersolid phase. The formation of these exotic quantum phases is mainly due to the combination effect and mutual competition of Raman-induced SOC, equilibrium  Rydberg interactions and immiscible contact interactions in the harmonic trapped BECs. In the scenario without external potential in Ref. \cite{Han}, each component density forms multiple uniformly distributed cells, as shown in Fig. 1 and Fig. 4 in Ref. \cite{Han}. Here the direct influence of the harmonic trap is that the total density profile of the two components tends towards the Thomas-Fermi distribution. At the same time the harmonic trap prevents the infinite free expansion of the uniform BECs, which can control the BECs and enhance the possibility for future experimental realization. Additionally, the external potential can stabilize quantum vortices for a rotating case, while the vortices in a uniform system are difficult to stabilize and hard to form easily observable topological defects. The structural differences and momentum distribution differences of these exotic quantum phases in Fig. 2 are expected to be tested and observed in the possible future cold-atom experiments.
\subsection{Destruction of chiral supersolid}
Next, we illustrate the effects of two commonly used regulation schemes in experiments, namely rotating the system and applying an in-plane quadrupole magnetic field, on the ground-state properties of the system. The results show that the two schemes break the CSS
phase of the system and lead to the formation of novel quantum phases. For the first case, we
consider a rotating system in a rotating frame by adding the term $-\Omega _{r}L_{z}\psi _{1}$($\psi _{2}$) to the right sides of the GP
equations (4) and (5) with the rotation frequency $\Omega _{r}$. Before the structural phase transition occurs (e.g., for relatively low rotation frequency $\Omega _{r}=0.1$), as shown in Fig. 3(c), we find that the density holes (corresponding to a visible vortex necklace) on the outermost layer of the density distribution in component 1 disappear, and the outer side of the second layer of the density distribution begins to break. At this point, the visible vortices in the system have not yet completely transformed into hidden vortices. However, the rapid rotation ($\Omega _{r}=0.8$) of the external potential breaks the original chiral supersolid structure (see Fig. 3(a), Fig. 1(b), and Fig. 2(VI)), where the original visible vortices outside the central region in component 1 disappear, and the density of component 1 exhibits periodic modulation along the azimuth direction, accompanied by the creation of double-layer hidden antivortex necklaces in both components. At the same time, the original helical antiskyrmion lattice in Fig. 1(b) is broken, and only one conventional antiskyrmion is generated at the center. Thus the ground state of the rotating system for large rotation frequency forms a special annular supersolid phase with double-layer hidden antivortex necklaces and a central Anderson-Toulouse coreless vortex. Our simulation results show that with the increase of the rotation frequency, the visible vortices in the system gradually evolve into hidden vortices from the outside to the inside, and the system undergoes a structural phase transition.

\begin{figure*}[htbp]
\centering
\centerline{\includegraphics*[width=15cm]{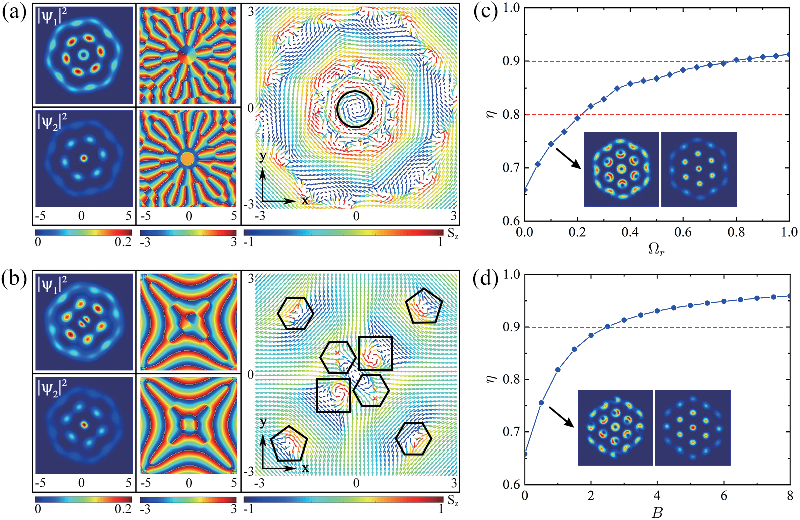}}
\caption{(Color online) (a) Ground-state density distribution, phase
distribution and spin texture of rotating two-component BECs with Raman-induced SOC and Rydberg interactions, where the rotation frequency $\Omega _{r}=0.8$, $\widetilde{C}_{6}=1500$, $\protect\kappa=4$, and $\Omega _{R}=10$. (b) Ground-state density distribution, phase
distribution and spin texture of non-rotating two-component BECs with Raman-induced SOC and Rydberg interactions in an in-plane quadrupole magnetic
field, where the quadrupole field strength $B=6$, $\widetilde{C}_{6}=1500$, $\protect\kappa=4$, and $\Omega _{R}=10$. The
arrows in the spin texture represent the transverse spin vector ($S_{x}$, $S_{y}$) and the color of each arrow indicates the magnitude of $S_{z}$. The square, circle, pentagon, and hexagon in the spin texture denote a skyrmion, an antiskyrmion, a half-skyrmion, and a half-antiskyrmion, respectively. Here each panel in (a) and (b) is a square, and the range and scale of the vertical axis are the same as those of the horizontal axis. The unit length in Figs. 3(a)-3(b) is $R_{c}$. (c) and (d) The miscibility $\protect\eta$ as a function of $\Omega _{r}$ and $B$ for $\widetilde{C}_{6}=1500$, $\protect\kappa=4$, and $\Omega _{R}=10$, respectively. The insets in Fig. 3(c) and Fig. 3(d) are the density distributions of the system for $\Omega _{r}=0.1$ and for $B=0.5$, respectively.}
\label{Fig3}
\end{figure*}

For the second case, we apply an in-plane quadrupole magnetic field (i.e., an in-plane gradient magnetic field) to the system by adding the term $g_{F}\mu _{B}\mathbf{B}\left( \mathbf{r}\right) \cdot \bm{\sigma }$ to the single-particle Hamiltonian \cite{Ray,Li}. Here $g_{F}=-1/2$ is the Lande factor, $\mu _{B}$ is Bohr magnetic moment, $\mathbf{B}\left( \mathbf{r}\right) =B\left(x\hat{e}_{x}-y\hat{e}_{y}\right) $ is the in-plane gradient magnetic field with the
strength $B$, and $\bm{\sigma }$ is the $2\times2$ Pauli spin matrices. In a weak in-plane quadrupole magnetic field (e.g., $B=0.5$), the density distribution of the system is similar to that at small rotation frequency (see Fig. 3(d)). The difference lies in that in this case, the fracture directions of the second layer density ring of component 1 are different, and the fracture occurs simultaneously on the inner and outer sides. Once the relatively strong in-plane quadrupole magnetic field is included, e.g., $B=6$ as shown in Fig. 3(b), the system exhibits an unusual density and phase distributions. Due to the presence of the in-plane quadrupole magnetic field, the density distributions of the two components are symmetrical about the two principal diagonals. In the meantime, the phase distributions of two components display typical quadrupole field characteristics, with hidden vortices and antivortices occupying the two principal diagonals, respectively. Particularly, the ground state of the system becomes a droplet lattice state with hidden vortices and antivortices along two principal diagonals. This ground-state structure is different from the case of spin-orbit-coupled dipolar BECs with in-plane quadrupole magnetic field (see Figs. 1(d) and 1(g) in Ref. \cite{Li}), where the ground state is an annular structure with hidden vortex-antivortex cluster in the central density hole region. In addition, our computation results show that the local topological charges in the spin texture in Fig. 3(b) approach $Q = 1$ (square marks), $Q = 0.5$ (pentagon marks), and $Q = -0.5$ (hexagon marks), which indicates that the local topological defects are skyrmion, half-skyrmion (meron), and half-antiskyrmion (antimeron), respectively \cite{Mermin}. Hence the spin texture forms a complex skyrmion-meron-antimeron cluster with the topological defects being distributed along the two principal diagonals.

Furthermore, the changes in the ground-state structure of the system for the above two cases can also be reflected by
the miscibility of the two components. A miscibility parameter $\eta$ can be defined to approximately measure the degree of overlap between the densities of the two components \cite{Kumar1,Kumar2},
\begin{equation}
\eta =2\int d\mathbf{r}\left\vert \psi _{1}\right\vert \left\vert \psi _{2}\right\vert
=2\int d\mathbf{r} \sqrt{\left\vert \psi _{1}\right\vert ^{2}\left\vert \psi
_{2}\right\vert ^{2}}.  \label{eta}
\end{equation}
This expression means that $\eta =1$ for the complete overlap between the two densities, and $\eta$ decreases as the overlap diminishes. In the absence of
rotation and gradient magnetic field, the CSS phase shows a clear spatial separation for the two components (see Fig. 1(b)), and the miscibility parameter $\eta$ has the smallest value as shown in Figs. 3(c) and 3(d). With the increase of the rotation frequency $\Omega _{r}$ or the quadrupole magnetic field strength $B$, the system gradually changes from an immiscible phase to a miscible phase in which $\eta \gtrsim 0.8$ \cite{Kumar1,Kumar2}. As the parameter $\Omega _{r}$ or $B$ further increases, the system becomes more miscible and eventually approaches a stable value. In particular, for the case of in-plane quadrupole magnetic field, the miscibility of the system achieves a greater value of $\eta >0.95$. This is due to fact that the strong magnetic field destroys the original visible vortex structure, resulting in the generation of hidden vortices and antivortices in both components, as well as the high overlap of some hidden vortices and antivortices in both components.
\subsection{Dissipative time crystal}
In Sec. III. A, we have discussed the ground-state properties of BECs with Raman-induced SOC and Rydberg interactions in a harmonic trap. Now we investigate the rotational dynamic behaviors of the system, where it is necessary to take into account dissipation. As a matter of fact, the dissipation in rotating BECs is universal and inevitable in actual cold-atom experiments. There are various ingredients that contribute to dissipation. For instance, the dissipation may be caused by the collision between the condensed atoms and the noncondensed atoms. At the same time, constant rotation drive or perturbation can trigger system heating and collective modes (such as surface mode and quadrupole mode) which attenuate and dissipate energy through nonlinear interactions \cite{Kasamatsu,WenLH}. In addition, the vortex motion in the rotating BECs (such as recombination, merging, or interaction with boundaries) may cause the system to emit phonons, leading to dissipation, and so on. Here we use a phenomenological dissipation model \cite{LWen,LWen2,Kasamatsu,WenLH} to study the rotating dynamics of the system. Moreover, Rydberg atoms are well controllable, which provides a suitable platform for the study of continuous time crystal (CTC). The CTC is a non-equilibrium quantum many-body state that spontaneously breaks time translation symmetry and exhibits stable periodic dynamics under continuous drive and dissipation, and it has an order parameter with self-sustained oscillations \cite{Kongkhambut,You,Watanabe,Krishna}. We are interested in exploring whether a similar CTC exists in this system. Therefore, the CTC in the rotating system can be revealed by examining the periodic (quasi-periodic) change of angular momentum with time and the periodic variation of density distribution. According to the phenomenological dissipation model \cite{LWen,LWen2,Kasamatsu,WenLH}, the nonlinear coupled GP equations (\ref{4})-(\ref{5})are transformed as follows
\begin{align}
\left( i-\lambda \right)\frac{\partial \psi _{1}}{\partial t}=& \Big[-\frac{1}{2}\nabla ^{2}+V(\mathbf{r
})+\beta _{11}|\psi _{1}|^{2}+\beta _{12}|\psi _{2}|^{2}  \notag \\
&+\frac{\Omega _{R}}{2}+\int U_{11}\left( \mathbf{r}-\mathbf{r}^{\prime }\right)
|\psi _{1}(\mathbf{r}^{\prime })|^{2}d\mathbf{r}^{\prime } \notag \\
&+\int U_{12}\left( \mathbf{r}-\mathbf{r}^{\prime }\right) |\psi _{2}(\mathbf{r}^{\prime })|^{2}d\mathbf{r}^{\prime }-\Omega_{r}L_{z}\Big]\psi _{1} \notag \\
& -\kappa \left(i\partial _{x}+\partial _{y}\right)\psi _{2}-\frac{\delta }{2}\psi _{2}, \label{dynamics1}
\end{align}
\begin{align}
\left( i-\lambda \right)\frac{\partial \psi _{2}}{\partial t}=& \Big[-\frac{1}{2}\nabla ^{2}+V(\mathbf{r
})+\beta _{22}|\psi _{2}|^{2}+\beta _{21}|\psi _{1}|^{2}  \notag \\
&-\frac{\Omega _{R}}{2}+\int U_{22}\left( \mathbf{r}-\mathbf{r}^{\prime }\right)
|\psi _{2}(\mathbf{r}^{\prime })|^{2}d\mathbf{r}^{\prime } \notag \\
&+\int U_{21}\left( \mathbf{r}-\mathbf{r}^{\prime }\right) |\psi _{1}(\mathbf{r}^{\prime })|^{2}d\mathbf{r}^{\prime }-\Omega_{r}L_{z}\Big]\psi _{2}  \notag \\
& -\kappa \left(i\partial _{x}-\partial _{y}\right)\psi _{1}-\frac{\delta }{2}\psi _{1}.  \label{dynamics2}
\end{align}
Here we choose the dissipation parameter $\lambda =0.005$, which corresponds
to a temperature of around $0.003T_{c}$ \cite{Kasamatsu}. A small dissipation parameter $\lambda $ slows the relaxation time to allow for detailed monitoring. But the variation of nonzero $\lambda $ does not change the dynamics of topological defect formation and the
ultimate steady structure of the rotating system. We select four typical
ground states as the initial states of dynamical evolution, and the specific parameters are (a) $
\widetilde{C}_{6}=250$, $\protect\kappa=4$, $\Omega _{R}=16$, (b) $\widetilde{C}_{6}=1500$, $\protect\kappa=4$, $\Omega _{R}=0$, (c) $
\widetilde{C}_{6}=1500$, $\protect\kappa=4$, $\Omega _{R}=10$, and (d) $\widetilde{C}_{6}=1500$, $\protect\kappa=7$, $\Omega _{R}=10$ for $\beta _{11}=\beta
_{22}=100$ and $\beta _{12}=200$. Note that the parameter values of (b), (c) and (d) are the same as those of (V), (VI) and (VII) in Fig. 2(c), respectively. Figs. 4(a)-4(d) show the temporal evolution of the mean orbital angular momentum per atom $\left\langle
L_{z}\right\rangle $ and typical density distribution of the system after
the harmonic trap begins to rotate suddenly with $\Omega _{r}=0.6$.

\begin{figure*}[tbhp]
\centering
\centerline{\includegraphics*[width=15cm]{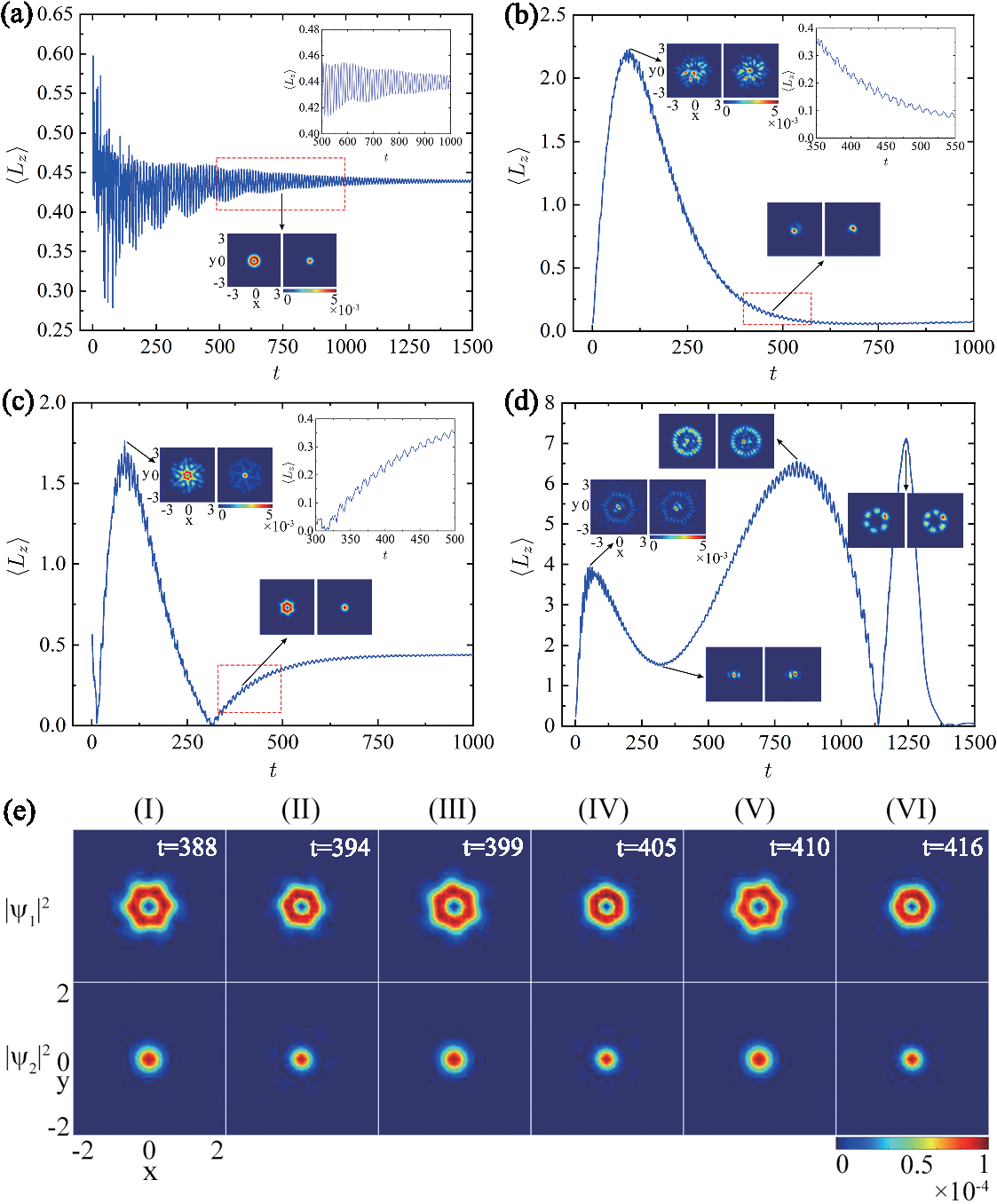}}
\caption{(Color online) (a)--(d) Temporal evolution of the average orbital
angular momentum per atom $\left\langle L_{z}\right\rangle$ for (a) $
\widetilde{C}_{6}=250$, $\protect\kappa=4$, $\Omega _{R}=16$, (b) $\widetilde{C}_{6}=1500$, $\protect\kappa=4$, $\Omega _{R}=0$, (c) $
\widetilde{C}_{6}=1500$, $\protect\kappa=4$, $\Omega _{R}=10$, and (d) $\widetilde{C}_{6}=1500$, $\protect\kappa=7$, $\Omega _{R}=10$. The insets in the upper right corners of panels (a)-(c) illustrate the local enlargements of the red dotted frames, respectively. The component density distributions at specific moments are shown in panels (a)-(d). (e) The temporal evolution of the density distribution corresponding to the red dotted frame in panel (c). Here $t$ and $\left\langle L_{z}\right\rangle$ are in
units of $\tau$ and $\hbar$, respectively. The unit length in Fig. 4(e) is $R_{c}$.}
\label{Fig4}
\end{figure*}

In Fig. 4(a), we find that the amplitude of $\left\langle L_{z}\right\rangle$ gradually changes
from an initial attenuation to a regular periodic persistent oscillation over time. This result is dramatically different from that of the spin-orbit-coupled BECs without soft-core long-range Rydberg interactions \cite{qWang}. For the latter case, the mean orbital angular momentum per atom increases rapidly with the time evolution and then gradually approaches a
maximum equilibrium value. The main reason for this oscillation is that the long-range interactions induced by Rydberg
dressing involve the coexistence and competition between multiple Rydberg states in the rotational dynamical
system \cite{You}. During the dynamical evolution, the system evolves from a multi-layer concentric toroidal stripe phase to an Anderson-Toulouse coreless vortex. However, when the system begins to rotate with a supersolid phase as the initial state,
the evolution of the orbital angular momentum over time is shown in Figs. 4(b)-4(d). If the SOC type in the system is Rashba
SOC as shown in Fig. 4(b), the results indicate that $\left\langle L_{z}\right\rangle $
first increases rapidly with the time evolution, then decreases in the form
of damping oscillation after reaching the peak, and finally oscillates with a
small amplitude near an equilibrium value.

For the system with Raman-induced SOC as displayed in Fig. 4(c), we find that a particular structural transition occurs at $t\approx 315$ before which $\left\langle L_{z}\right\rangle $ experiences irregular quasi-periodic alternating complex oscillations composed of downward damping oscillations and upward amplification oscillations. Then $\left\langle L_{z}\right\rangle $ increases in a quasi-periodic oscillation
pattern until the oscillation is stable. The quasi-periodic oscillation in the process of dynamical evolution is strong evidence of the time crystal. Fig. 4(e) shows the typical dynamics of the density distribution, where $\widetilde{C}_{6}=1500$, $%
\kappa=4$, $\Omega _{R}=10$, and the representative time points are (%
\uppercase\expandafter{\romannumeral1}) $t=388$, (\uppercase%
\expandafter{\romannumeral2}) $t=394$, (\uppercase\expandafter{%
\romannumeral3}) $t=399$, (\uppercase\expandafter{\romannumeral4}) $t=405$, (%
\uppercase\expandafter{\romannumeral5}) $t=410$, and (\uppercase%
\expandafter{\romannumeral6}) $t=416$. Our results indicate that the distribution area of the
condensate varies periodically with time, and the condensate area achieves the maximum
at the trough of $\left\langle L_{z}\right\rangle$ oscillation and the minimum at the crest of $\left\langle L_{z}\right\rangle$ oscillation. As is well known, if a quantum state is an eigenstate of the angular momentum operator $L_{z}$, the expected values $\left\langle L_{x}\right\rangle$ and $\left\langle L_{y}\right\rangle$ of the angular momentum operators $L_{x}$ and $L_{y}$ will both be zero. Obviously, the system spinor wave function during the dynamical evolution is not the eigenstate of the angular momentum operator $L_{z}$, therefore the expected values $\left\langle L_{x}\right\rangle$ and $\left\langle L_{y}\right\rangle$ are non-zero. The dissipation considered here is weak, so its influence on the angular momentum in a short time can be ignored. Physically, for a constant rotation frequency, when the $z$ component of the angular
momentum decreases, the $x$ and $y$ components of the angular momentum
increase, and vice versa. $\left\langle L_{z}\right\rangle$ is small at the trough and large at the crest, which means that $\left\langle
L_{x}\right\rangle$ and $\left\langle L_{y}\right\rangle$ are large at the trough and small at the crest. This point can explain the periodic change of the condensate area in the $x$-$y$ plane. The continuous periodic changes of the
condensate are significant characteristics of CTC. Here the density
distribution of the CTC turns out to be a phase separation structure and the chiral symmetry of the system is broken. Thus the appearance of quasi-periodic oscillation part of the system indicates the existence of chiral CTC. In Fig. 4(d), the results show that the system has
multiple transition points due to the larger SOC strength and the more
complex density distribution of the SWSS phase in the selected initial
state. Moreover, the variation of the angular momentum input is affected by
the complex competition among nonlinear contact interactions, SOC, Rydberg interactions and rotation during the process of dynamical evolution.
\section{Conclusions}
In summary, we have investigated the ground-state properties and rotational dynamic behaviors of quasi-2D
two-component BECs with Raman-induced SOC and Rydberg interactions. We show that the Raman-induced SOC can lead to the formation of a novel chiral supersolid phase with a helical antiskyrmion lattice in two-component BECs with balanced Rydberg interactions in a harmonic trap. We present two ground-state phase diagrams in which one is spanned by the Rydberg interaction strength and the SOC strength, and  another is with respect to the Rydberg interaction strength and the Raman coupling strength. This system exhibits a rich variety of exotic quantum phases, including half-quantum vortex phase, stripe supersolid phase, chiral supersolid phase, standing-wave supersolid phase, toroidal stripe phase with a central Anderson-Toulouse vortex, mirror-symmetric supersolid phase, and checkerboard supersolid phase. These quantum phases can be achieved by adjusting the Rydberg interaction strength, the Raman coupling strength, or the SOC strength. Particularly, the latter three quantum phases have not been reported elsewhere. In addition, both rotation and in-plane quadrupole magnetic field can break the chiral supersolid phase and make the ground-state structure develop towards a miscible phase. Furthermore, we have discussed the rotating dynamics of Rydberg-dressed BECs with Raman-induced SOC in a harmonic trap by using a phenomenological dissipation model. We find that when the initial state is a chiral supersolid phase the rotating system supports dissipative continuous time crystals. This system is theoretically feasible and can be attained in principle. With the on-going development of cold-atom experimental techniques, the system may be achieved in the future and these novel quantum phases and dynamic properties are expected to be tested and observed in experiments. These findings in the present work enrich our new knowledge and insight for the peculiar matter states such as supersolid, superfluid, and time crystal in cold atom physics and condensed matter physics.

\begin{acknowledgments}
This work was supported by the National Natural Science Foundation of China (Grant Nos. 11475144 and 11047033), Hebei Natural Science Foundation (Grant Nos. A2022203001, A2019203049 and A2015203037), Research Foundation of Yanshan University (Grant No. B846), and Double First-Class Team Foundation of Yanshan University.
\end{acknowledgments}

\end{document}